\documentclass[12pt]{article}

\usepackage{scicite}
\usepackage{graphics}
\usepackage{times}
\usepackage{siunitx}
\usepackage{float}
\usepackage{xr}
\usepackage{bm}
\usepackage{authblk}
\usepackage{afterpage}

\usepackage{graphicx}
\usepackage[labelfont=bf]{caption}

\newenvironment{sciabstract}{%
\begin{quote} \bf}
{\end{quote}}

\newcounter{lastnote}

\title{Directly observing atomic-scale relaxations of a glass forming liquid using femtosecond X-ray photon correlation spectroscopy}

\author{ \small
Tomoki Fujita$^1$, Yanwen Sun$^{2\ast}$, Haoyuan Li$^3$, Thies J. Albert$^4$, Sanghoon Song$^2$, Takahiro Sato$^2$, Jens Moesgaard$^1$, Antoine Cornet$^5$, Peihao Sun$^6$, Ying Chen$^2$, Mianzhen Mo$^2$, Narges Amini$^1$, Fan Yang$^7$, Arune Makareviciute$^1$, Garrett Coleman$^8$, Pierre Lucas$^8$, Jan Peter Embs$^9$, Vincent Esposito$^2$, Joan Vila-Comamala$^{9}$, Nan Wang$^2$,Talgat Mamyrbayev$^{9}$, Christian David$^{9}$, Jerome Hastings$^2$, Beatrice Ruta$^5$, Paul Fuoss$^2$, Klaus Sokolowski-Tinten$^4$, Diling Zhu$^2$, Shuai Wei$^{1,10\ast}$ 
\\
\normalsize{\small $^{1}$Department of Chemistry, Aarhus University, 8000 Aarhus, Denmark} \\
\normalsize{\small $^{2}$Linac Coherent Light Source, SLAC National Accelerator Laboratory, Menlo Park, CA 94025, USA} \\
\normalsize{\small $^{3}$Department of Mechanical Engineering, Stanford University, 440 Escondido Mall, Stanford, CA 94305, USA} \\
\normalsize{\small $^{4}$Faculty of Physics and Center for Nanointegration Duisburg-Essen (CENIDE), University of Duisburg-Essen, Lotharstrasse 1, 47048 Duisburg, Germany} \\
\normalsize{\small $^{5}$Institut Néel, CNRS, 38042 Grenoble, France} \\
\normalsize{\small $^{6}$Department of Physics, Università degli Studi di Padova, 35122 Padova, Italy}\\
\normalsize{\small $^{7}$Institute of Materials Physics, Deutsches Zentrum für Luft- und Raumfahrt (DLR), Köln, Germany}\\
\normalsize{\small $^{8}$Department of Materials Science and Engineering, University of Arizona, Tucson AZ 85721, USA}\\
\normalsize{\small $^{9}$Paul Scherrer Institut, 5232 Villigen PSI, Switzerland}\\
\normalsize{\small $^{10}$Centre for Integrated Materials Research, Aarhus University, 8000 Aarhus, Denmark} \\
\normalsize{\small $^\ast$To whom correspondence should be addressed; E-mail: yanwen@slac.stanford.edu, shuai.wei@chem.au.dk.}
}

\DeclareUnicodeCharacter{2061}{}

\date{}

\begin{document} 


\baselineskip24pt


\maketitle

\begin{sciabstract}
Glass forming liquids exhibit structural relaxation behaviors, reflecting underlying atomic rearrangements on a wide range of timescales. These behaviors play a crucial role in determining many material properties. However, the relaxation processes on the atomic scale are not well understood due to the experimental difficulties in directly characterizing the evolving correlations of atomic order in disordered systems. Here, taking the model system $\bm{\mathrm{Ge_{15}Te_{85}}}$, we demonstrate an experimental approach that probes the relaxation dynamics by scattering the coherent X-ray pulses with femtosecond duration produced by X-ray free electron lasers (XFELs). By collecting the summed speckle patterns from two rapidly successive, nearly identical X-ray pulses generated using a split-delay system, we can extract the contrast decay of speckle patterns originating from sample dynamics and observe the full decorrelation of local order on the sub-picosecond timescale. This provides the direct atomic-level evidence of fragile liquid behavior of $\bm{\mathrm{Ge_{15}Te_{85}}}$. Our results demonstrate the strategy for XFEL-based X-ray photon correlation spectroscopy (XPCS), attaining femtosecond temporal and atomic-scale spatial resolutions. This twelve orders of magnitude extension from the millisecond regime of synchrotron-based XPCS opens a new avenue of experimental studies of relaxation dynamics in liquids, glasses, and other highly disordered systems.
\end{sciabstract}

\section{Introduction}
Liquids and glasses exhibit structural relaxations of atomic rearrangements toward their energetically favorable positions. The relaxation dynamics are extremely diverse with the characteristic timescales ranging from millions of years for glass ageing to some sub-picoseconds for fast relaxation processes in high-fluidity liquids. They are critical to many properties of glass forming systems, such as viscosity, vitrification, amorphous stability, and crystallization~\cite{wang_dynamic_2019,angell_relaxation_2000,tanaka_roles_2022, ruta_relaxation_2017, dyre_colloquium_2006}. Relaxation dynamics are usually characterized by measuring  time (or frequency)-dependent changes of macroscopic properties using techniques such as dynamic mechanical spectroscopy, dielectric spectroscopy, calorimetry and rheology~\cite{ruta_relaxation_2017, monnier_vitrification_2020, cheng_highly_2022, weeks2000three, richert2014supercooled}.
However, the atomic-level mechanisms of structural relaxations are often debated, as few experimental techniques can directly probe the atomic-scale structural relaxations in disordered systems. Over the last two decades, X-ray photon correlation spectroscopy (XPCS) based on synchrotron X-ray sources has been developed to determine the intensity autocorrelation functions from measured speckle patterns~\cite{ruta_atomic-scale_2012, shpyrko2014x, ruta_wave-vector_2020}, and thus determining the intermediate scattering functions (ISF) and revealing the relaxation dynamics on the atomic length scales~\cite{shpyrko2014x}. However, XPCS has been limited to slow dynamics near and below the glass transition temperature $\mathrm{T_g}$, and it has been challenging to make measurements on the microsecond timescale or below due to the limited coherent photon flux at synchrotron X-ray sources~\cite{grubel2007xpcs}.

Direct experimental access to faster relaxation dynamics is of particular interest, because numerous glass forming liquids exhibit short relaxation times (i.e., high fluidity) at high temperatures due to their high liquid fragility. The fragility concept classifies the diverse variety of liquids according to their temperature dependence of relaxation times (or viscosity)~\cite{angell1995formation}. On approaching $\mathrm{T_g}$, some liquids exhibit a near-Arrhenius rise in viscosity, classified as ``strong", while others, as ``fragile liquids", show a range of super-Arrhenius behavior~\cite{angell1995formation}. While many liquids follow a simple fragile or strong behavior, some anomalous liquids exhibit a so-called fragile-to-strong transition (FST) (also referred as dynamic crossover)~\cite{angell2002liquid, tanaka2020liquid,wei2022anomalous}. A FST is usually accompanied by thermodynamic response function maxima (e.g. heat capacity and compressibility), as well as local structural changes~\cite{wei2015phase,zalden2019femtosecond,wei2019phase}. Such a transition has been long debated in water~\cite{angell2002liquid, gallo2016water, kim2017maxima, dehaoui2015viscosity} and suggested in silicon~\cite{sastry2003liquid, jakse2007liquid, zhao2016phase, vasisht2011liquid}, germanium~\cite{bhat2007vitrification}, oxides, molecular and metallic systems, and many others~\cite{tanaka2020liquid,henry2020liquid,xu2006relationship,stolpe2016structural, wei2022anomalous}. A clear FST has been demonstrated in liquid $\mathrm{Ge_{15}Te_{85}}$ near its eutectic melting point~\cite{wei2015phase}. In the related systems (e.g. $\mathrm{Ge_{15}Sb_{85}}$ and $\mathrm{AgInSbTe}$), the FST plays an important role in the functionality of phase-change memory devices~\cite{zalden2019femtosecond, wei2019phase, orava2015fragile}. Yet, the atomic-level understanding of relaxation dynamics near these transitions presents a tremendous challenge, because 1) FSTs reported in literature typically occur far above $\mathrm{T_g}$ with a short relaxation timescale from nanoseconds to sub-picoseconds, far beyond the capability of synchrotron-based XPCS, and 2) FSTs are often hidden in the supercooled liquid below the melting temperature $\mathrm{T_m}$ obscured by fast crystallization.

X-ray free electron lasers (XFELs), delivering unprecedentedly high numbers of photons within sub-100-femtosecond pulses with nearly full transverse coherence, present the opportunity for developing the XPCS techniques capable of probing dynamics in the regime from femtosecond to nanosecond timescales. The key idea is that a femtosecond X-ray pulse is split into two nearly identical coherent pulses with an adjustable time delay in-between. As the double pulses scatter from the sample in rapid succession, the summed speckle patterns, collected by a 2D detector, carry information about atomic dynamics on the timescale of the double pulse separation. However, implementing this concept has faced technical challenges including generating identical double pulses~\cite{sun2020realizing}, extremely low count rates (limited at wide angles due to the small scattering cross section from atomic scale order) of order  $10^{-4}$ photons/pixel~\cite{hruszkewycz2012high}, and lack of robust analytical methods of extracting dynamics from noise or artifacts.

Pioneering works have explored the feasibility of split-delay optics to deliver double X-ray pulses with identical properties~\cite{roseker2009performance,osaka2013bragg, osaka2016wavelength, zhu2017development,lu2018development, rysov2019compact, sun2019compact}, required for extracting the dynamics information of the sample. 
The conventional division-of-wavefront split-delay line, despite providing double pulses with good efficiency, suffers from the instability of the crystal-optics-based beam splitters and has difficulty in preserving sufficient mutual coherence between the two pulses~\cite{sun2020realizing}. 
A more recent approach, introduced by some of the authors of this work, employs an amplitude-splitting delay line using transmission grating-based splitters and has generated highly mutually coherent hard X-ray pulse pairs~\cite{li2021generation}. 
In addition, the femto- to pico-seconds separation of pulse pairs is beyond the time resolution of any X-ray detectors, which renders the data analysis strategy significantly different from that of synchrotron-based XPCS. Contrast extraction, relying on analyzing photon statistics of the summed speckles~\cite{hruszkewycz2012high}, has been developed to obtain the ISF for nanometer-scale dynamics from measurements at small angles~\cite{roseker2018towards,sun2021nonuniform}. However, access to atomic-scale dynamics has been difficult due to the orders of magnitude lower scattering signals at wide angles. Concerns revolve around two key issues: 1) the double pulses might fail to maintain a high level of mutual stability during the extended period of data acquisition; 2) detector noise, artifacts, and background radiation might interfere with the accurate extraction of speckle contrast.
A recent study of water observed the speckle contrast decay with a split-delay line, possibly reflecting water's atomic-scale dynamics; yet, it is unclear whether the double-pulse overlap remained stable during the measurement and any drift in overlap would lead to errors in the measured dynamics~\cite{shinohara2020split}.

The glass forming system Ge$_{15}$Te$_{85}$ is known to undergo a dynamic crossover at around 400~$\mathrm{^{\circ}C}$, where the viscosity drops sharply by two orders of magnitude to a high-fluidity fragile state. However, the atomic-level dynamics in this state have not been observed, making Ge$_{15}$Te$_{85}$ an excellent initial system on which to develop and demonstrate the capabilities of XPCS with split XFEL pulses. 

In this work, we probe the fragile state by employing a XFEL-based XPCS technique with femtosecond time resolution (fs-XPCS) using the aforementioned amplitude splitting delay line~\cite{li2021generation} at the Linac Coherent Light Source (LCLS) (XPP beamline) at SLAC. We develop a method of simultaneously measuring the small-angle scattering (SAXS) and wide-angle scattering (WAXS) signals to account for the spatial overlap of the pulse pairs. We show that even with extremely low photon counts, contrast extraction with photon statistics analysis allows for observation of speckle contrast decay resulting from the sample dynamics. We demonstrate a strategy to determine the ISF on the sub-picosecond timescale, providing the direct atomic-level evidence of fast dynamics in $\mathrm{Ge_{15}Te_{85}}$ after it is transformed to a fragile liquid.

\section{Results}
 \begin{figure}[H]
    \centering
    \includegraphics[width=\textwidth]{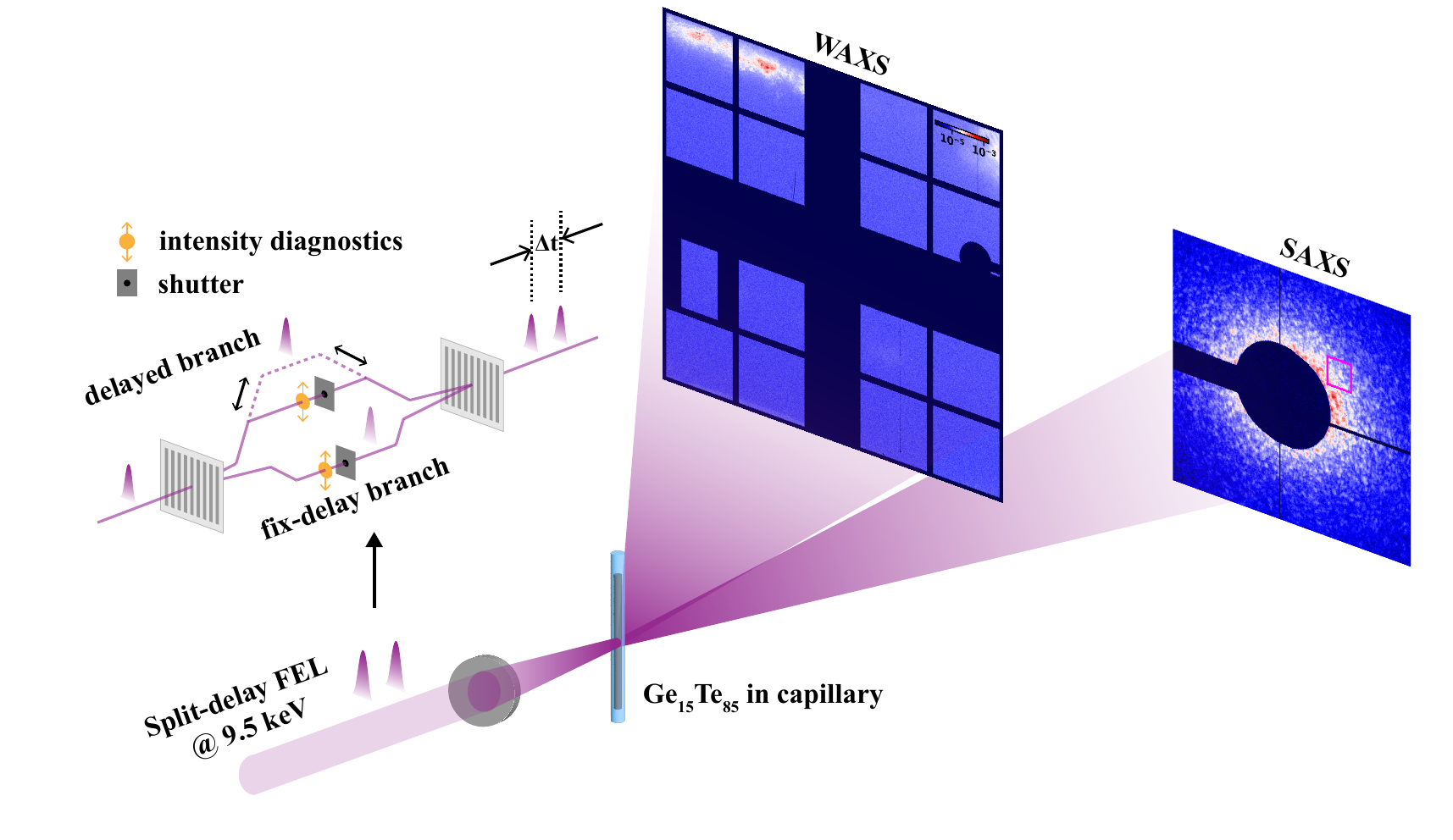}
    \caption{\textbf{The schematic of the experimental setup of the femtosecond X-ray photon correlation spectroscopy (fs-XPCS).} The grating-based split-delay optics are schematically shown on the left panel. An array of WAXS and SAXS detectors, placed downstream of the sample, collects speckle patterns at wide and small angles simultaneously. Here we show the WAXS and SAXS patterns averaged over 1 million and 100 shots respectively.}
    \label{fig:schematics}
\end{figure}
Figure~\ref{fig:schematics} shows the experimental schematic of the fs-XPCS setup.
 The output pulse pairs from the split-delay system~\cite{li2021generation} with a photon energy of 9.5~keV are focused on the sample location by a set of compound refractive beryllium lenses with a focal length of 1.5~m. The sample,  $\mathrm{Ge_{15}Te_{85}}$, is encapsulated in a quartz capillary with an inner diameter of about 10~$\mu$m, and resistively heated to the fragile liquid state at 550~$\mathrm{^{\circ}C}$ (well above the eutectic melting point of 385~$\mathrm{^{\circ}C}$). The sample temperature is constantly monitored by two thermocouples placed close to the sample position on both sides of the capillary. The temperature stability is within 0.1~$\mathrm{^{\circ}C}$ throughout the measurements and the difference between two thermocouples stays within 4~$\mathrm{^{\circ}C}$. The focal spot size is approximately 2~$\mu$m (FWHM) and the average pulse pair energy is characterized to be 0.15~$\mu$J at the sample plane. Four ePix100 detectors~\cite{sikorski2016application} are assembled in a 2 by 2 array and placed 2.5~m downstream to cover the first structure factor $S(\mathrm{Q})$ peak (at Q$_0=2.0~\mathrm{\AA^{-1}}$) of the sample to probe its atomic-scale relaxation dynamics. Another ePix100 detector is mounted 5~m downstream to simultaneously measure the SAXS signals, which come mostly from the quartz capillary and allow for \textit{in-situ} characterization of the spatial overlap of the two pulses. 

Ensuring the stable and highly overlapped condition of pulse pairs is essential for distinguishing intrinsic sample dynamics from many possible instabilities and artifacts. In Fig.~\ref{fig:schematics} (left panel), each XFEL pulse is split into two pulses via a diamond transmission grating~\cite{li2021generation}. 
The two pulses are directed by Bragg crystal reflections to travel along different optical paths with the path length difference determining their time delay $\Delta t$. They are then recombined to the same trajectory by the other diamond grating further downstream. Transmissive intensity diagnostics consisting of a diode collecting the scattering signal from a Kapton target are placed in the optical paths of individual branches to measure their intensities on a shot-to-shot basis. One shot consists of a single pulse or pulse pairs depending on the measurement mode. 
By installing shutters in the respective optical paths for the two beams, we can constantly cycle between three modes of pulses: 1) the single pulse per shot through the path length adjustable branch (i.e. delayed branch), 
2) the single pulse per shot through the fixed path branch (i.e. fix-delay branch), and 3) the pulse pairs per shot through both branches. As a result, speckle patterns for each shot were collected as ``data frames" at both SAXS and WAXS detectors. Each data frame corresponds to one of the three modes (i.e., delayed, fix-delay, both).  
By comparing the speckle contrast of SAXS signals from the three measurements, we are able to monitor the degree of transverse coherence as well as the spatial overlap of the two pulses continuously. 
This information provides the feedback to the analysis of the wide-angle contrast degradation, mandatory for the accurate extraction of the ISF for studying sample dynamics as discussed below.

\begin{figure}[h!]
    \centering
    \includegraphics[width=0.95\textwidth]{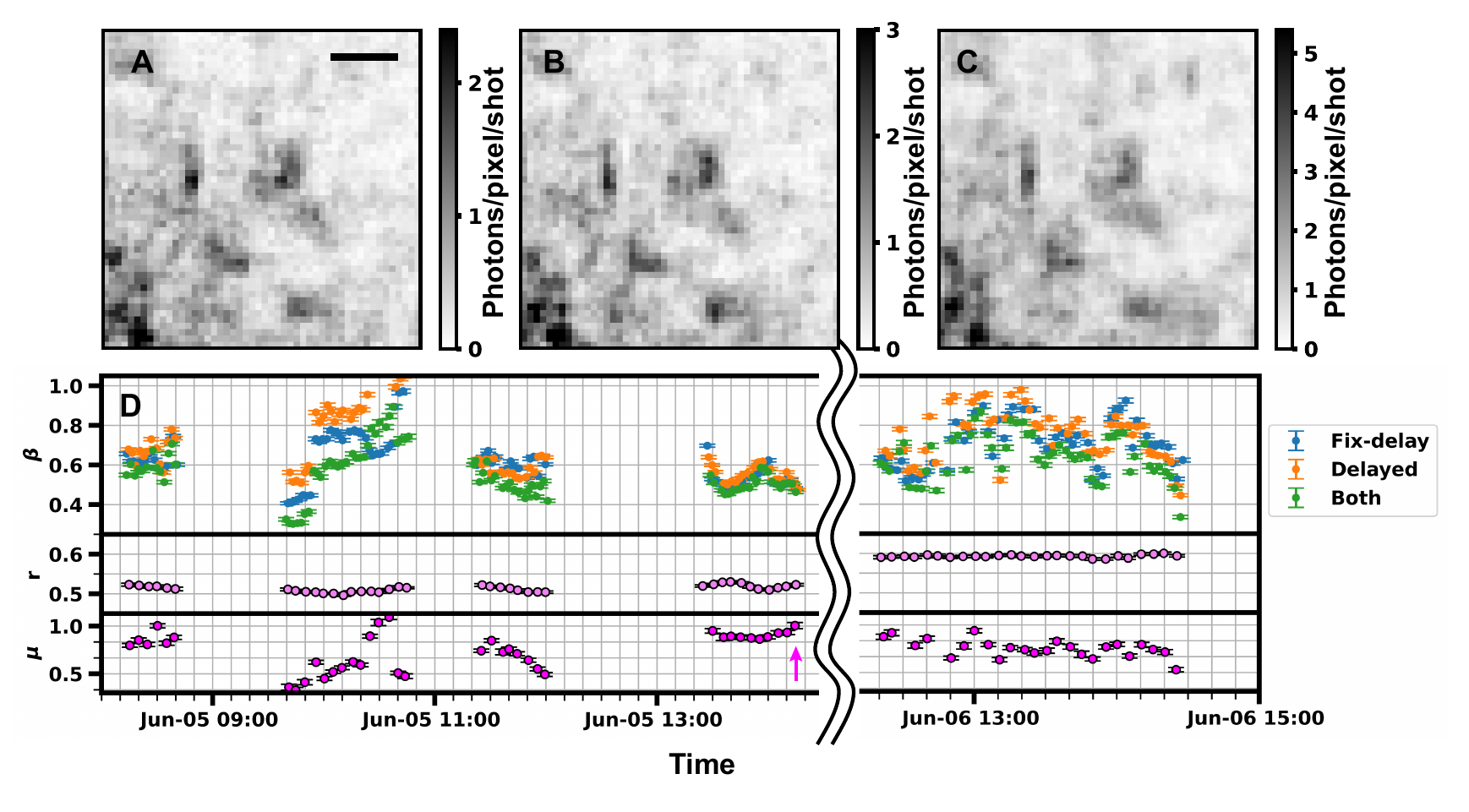}
    \caption{\textbf{The small-angle coherent scattering from the static reference for obtaining the real-time overlap coefficient of the pulse pairs.} The speckle patterns are averaged over 40 frames for the delayed (\textbf{A}), fix-delay (\textbf{B}) and both (\textbf{C}) branch modes. They are measured, while collecting the wide-angle data, at the delay time of 0.1~ps. The displayed region corresponds to the square highlighted in magenta in the SAXS detector shown in Fig.~\ref{fig:schematics}. The gray scales show the count rate and have a unit of photons/pixel/shot. The scale bar in (\textbf{A}) is the same for all 3 plots and corresponds to 0.0005 $\mathrm{\AA^{-1}}$.  (\textbf{D}) The speckle contrasts $\beta$ (upper panel), the branching ratio $r$ of the pulse pairs, and the overlap coefficient $\mu$ (lower panel) obtained from the intensity diagnostic and the SAXS data with the elapse of the measurement time. During the same period, the wide-angle speckle data for 0.1~ps and 0.65~ps delays are collected simultaneously. The magenta arrow points to the time when the speckles in (\textbf{A}-\textbf{C}) are measured.}
    \label{fig:SAXS}
\end{figure}

The contrast of the summed speckle pattern from both beams ($\beta_\mathrm{both}$) can be related to that of the single branch beams ($\beta_\mathrm{fix-delay}$ and $\beta_\mathrm{delayed}$) with~\cite{sun2021nonuniform}
\begin{equation}
   \beta_\mathrm{both} = r^2 \beta_\mathrm{fix-delay} + (1-r)^2 \beta_\mathrm{delayed}+ r(1-r)(\beta_\mathrm{fix-delay}+ \beta_\mathrm{delayed}) \mu F,
    \label{eq:ISF}
\end{equation}
where $F = |f|^2$ is an experimentally accessible quantity by fs-XPCS and $f=f(\mathrm{Q}, \Delta t)$ is the ISF to be determined for characterizing the relaxation dynamics, at a given scattering wavevector Q and a time delay $\Delta t$. The branching ratio $r = i_\mathrm{fix-delay}/(i_\mathrm{fix-delay}+i_\mathrm{delayed})$ is the intensity fraction of the fix-delay branch in the pulse pair (see Supplementary Material Section~1 for details regarding the intensity distribution of the split-delay system). 
The beam overlap coefficient $\mu$ ($0\leq \mu \leq 1$) characterizes the effect due to the imperfect spatial beam overlap between the two branches~\cite{roseker2018towards}. $\mu = 1$ corresponds to an ideal overlap, while in the case of $\mu = 0$, the two beams are probing different sample locations, and the contrast from their summed speckle pattern $\beta_\mathrm{both}$ no longer encodes sample dynamics, i.e., ISF. We stress that since both $\mu$ and ISF contribute to the change in $\beta_\mathrm{both}$, extracting the contrast changes alone does not yield the ISF. 
As a result, a sufficiently high overlap level and a real-time diagnostic of the overlap coefficient are a prerequisite for accurately extracting the ISF. 

In the experiment, the SAXS intensity serves as a static reference, i.e., $|f(\mathrm{Q}, \Delta t)| = 1$, as the observation timescale (of order $\sim$ps) is far smaller than the relaxation time of the quartz glass (i.e. fused silica) at this low-Q region at a temperature well below its $\mathrm{T_g} = 1173 \mathrm{^{\circ}C}$. We are therefore able to quantify $\mu$ directly from the single and both beam SAXS contrasts.
Figure~\ref{fig:SAXS} (\textbf{A}-\textbf{C}) show a region in the averaged speckle patterns of the SAXS data corresponding to the three measurement modes. 
The high similarity of the three patterns indicate an excellent spatial overlap between the two pulses. Their contrasts as a function of the measurement time are shown in the upper panel of Fig.~\ref{fig:SAXS} (\textbf{D}). 
Since the transmissive intensity diagnostic provides a real time measurement of the branching ratio as shown in the middle panel, the overlap coefficient $\mu$ can be derived using Equation~\ref{eq:ISF} and is shown in the lower panel. 
It is clear that $\mu$ drifts on a minute timescale. The value of $\mu$ remains mostly above 0.6, indicating a satisfactory overlap, thanks to the real-time monitoring and re-alignment of the beams, although $\mu$ of some frames may occasionally drift well below 0.6 (e.g. on Jun-05 9:00-9:30 as in Fig.~2 (\textbf{D})). For the consistency of analysis, we have excluded the WAXS data frames with $\mu<0.6$. Furthermore, an effective overlap value $\bar{\mu}$ is derived from the averaged SAXS contrast values of the three modes during the exact same time span of the WAXS measurements. This value is then inserted to Equation 1 to benchmark the WAXS contrast reduction and compute $F$ for each time delay, as detailed in Table~\ref{table:photon_statistics}. 
\begin{table}[h!]
    \centering
    \resizebox{\columnwidth}{!}{
    \begin{tabular}{|c|c|c|c|c|c|c|c|c|c|}
        \hline
        $\Delta t$ (ps) & $\bar{k}$ (photons/pixel) & $N_\mathrm{frame}$ &  $\hat{\beta}$ &  $\delta \beta_\mathrm{MLE}$& $\delta \beta_\mathrm{Poisson}$ & $\bar{\mu}$ & $\bar{r}$ & $F$\\
         \hline
        0.1 &  5.8 $\times 10^{-5}$ & 257250 &  0.17& 0.05 & 0.05 & 0.80 (0.03) & 0.51 (0.05) & 1.3 (0.9)\\
         \hline
         0.65  & 4.6 $\times 10^{-5}$ & 283067 &  0.04& 0.05 & 0.05 & 0.76 (0.08) & 0.59 (0.07)& -0.7 (0.8)\\
         \hline
    \end{tabular}
    }
    \caption{Measurement statistics, parameters and results including the count rate $\bar{k}$, the number of frames $N_\mathrm{frame}$ and pixels $N_\mathrm{pixel} \approx 7\times 10^5$, the speckle contrast levels $\hat{\beta}$, their error bars from both the maximum likelihood fitting $\delta \beta_\mathrm{MLE}$ and the photon statistics $\delta \beta_\mathrm{Poisson}$ for the two delays, i.e., 0.1 ps and 0.65 ps. The effective overlap $\bar{\mu}$ from the SAXS monitoring and the averaged branching ratio $\bar{r}$ from the intensity diagnostic, corresponding to the same period of WAXS data collection, are used to calculate $F$. The error range is reported in the parentheses.} 
    \label{table:photon_statistics}
\end{table}

The wide-angle speckle patterns collected at the first $\mathrm{S(Q)}$ maximum contain the information of the atomic-scale characteristic relaxation dynamics of the fragile liquid state. Since the scattering cross-section from the atomic-scale order is limited, the count rate is on average about $5\times 10^{-5}$ photons/pixel per shot with both beams illuminating the sample. 
It therefore requires a large number of data frames from detectors containing multi-million pixels and the careful treatment of artifacts including abnormal detector pixels, background radiation from impurities in concrete and cosmic rays~\cite{baron2023background}, etc, to accurately extract the contrast values from photon statistics, i.e., from the probability of multiple photon per pixel events (see Supplementary Material Section~2). 
The method of obtaining contrast from speckle patterns with discrete photon events has been demonstrated by Hruszkewycz et, al in~\cite{hruszkewycz2012high}, albeit at a much larger count rate, on average more than 0.01 photons/pixel. At this count rate, beam-induced permanent changes are visible when the sample was evaluated with a scanning electron microscope. 
It remains experimentally unclear whether the contrast extraction method can be further applied to even lower count rates in the non-perturbed regime, by simply extending the measurement time.  
Moreover, since monochromatic SASE (self-amplified spontaneous emission) pulses from an XFEL exhibits large intensity fluctuations~\cite{zhu2014performance}, it is essential to appropriately weigh the signal contribution of individual frames to the overall photon statistics. Therefore, we adopt a different approach to obtain the contrast by introducing a maximum likelihood based contrast estimator (MLE). The likelihood ratio $\chi^2$ is defined as~\cite{roseker2018towards, wilks1938large}
\begin{equation}
    \chi^2 (\beta) = -2\sum_{f = 1}^{N_\mathrm{frame}} \sum_{k = 0} ^ {n_k}   p_{f,k} N_\mathrm{pixel} \ln (  \frac{P_k(\beta, \bar{k_f})}{p_{f,k}})
    \label{equation: chi2}
\end{equation}
For each frame $f$ with available pixels $N_\mathrm{pixel}$, the probabilities of each pixel having photons $k$ = 0, 1, 2, denoted as $p_{f, k}$, are extracted from the WAXS data (we only consider up to $n_k = 2$~photon/pixel events).  $p_{f, k}$ are then compared with the projected theoretical probability $P_k$ (using Equation 1 in Supplementary Material) for a given count rate $\bar{k_f}$. 
The optimal estimate of the contrast, $\hat{\beta}$, is obtained by finding the minimum $\chi^2$ values from the numerical calculations. The error $\delta \beta_{\mathrm{MLE}}$ can be retrieved by computing the second derivative of $\chi^2$ at $\hat{\beta}$ 
\begin{equation}
    \delta^2 \beta_{\mathrm{MLE}} = ( \frac{ \partial^2 \chi^2 (\beta)}{2!\partial \beta^2})^{-1}|_{\beta = \hat{\beta}}.
    \label{equation: chi2_error}
\end{equation}

\begin{figure}[h!]
    \centering
    \includegraphics[width=0.65\textwidth]{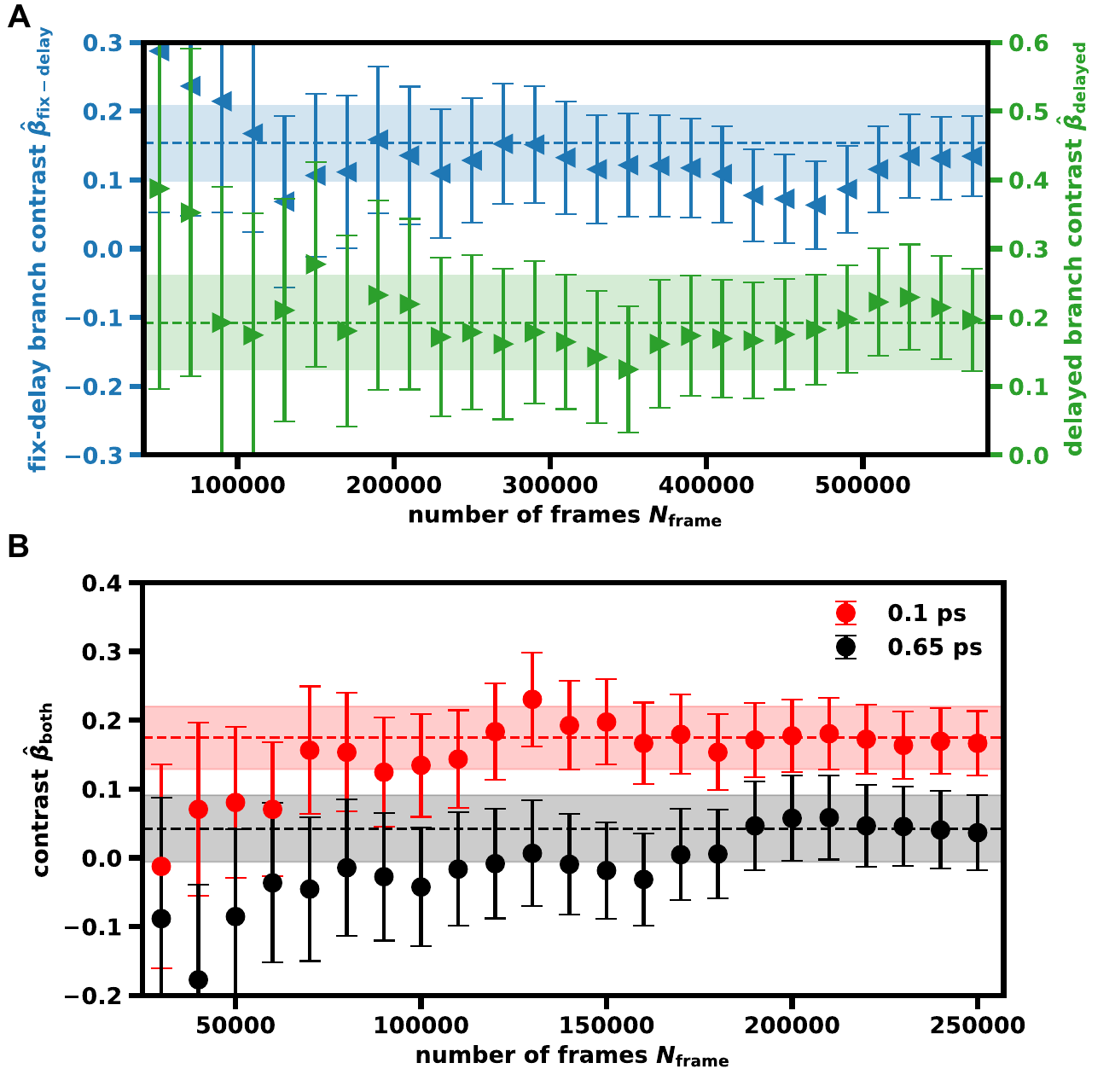}
    \caption{
    \textbf{Convergence of contrasts for (\textbf{A}) single-pulse and (\textbf{B}) both-beam mode.} For each beam condition, the dashed line and the shaded area show the final contrast value and the MLE error when using all the data, which correspond to Supplementary Material Fig.~3. The contrasts for the fix-delay (blue) and the delayed branch (green) in (\textbf{A}) are offset by using double y-axes for clarity. For the both-beam mode in (\textbf{B}), the clear difference in final contrast values between 0.1 and 0.65 ps  indicates a contrast decay.
    }
    \label{fig:fig3}
\end{figure}

This method is used to determine the contrast values for each mode of WAXS measurements, including the two single-pulse modes, (i.e., fix-delay and delayed branch) as well as the both-beam mode at 0.1~ps and 0.65~ps, as shown in Fig.~3 of the Supplementary Material. We furthermore plot the results of the contrast extraction, which slowly converge with increasing number of frames (see Fig.~\ref{fig:fig3} (\textbf{A}, \textbf{B})). The error bars are estimated using Equation~\ref{equation: chi2_error}. For the two delays, i.e., 0.1~ps and 0.65~ps, the difference in their contrast becomes significantly larger than their errors when accumulating over 100000 shots. Adding more frames further reduces the error, with the final values summarized in Table~\ref{table:photon_statistics}. 
Strikingly, the MLE error $\delta \beta_{\mathrm{MLE}}$ is in an excellent agreement with the error derived from the photon counting statistics,
\begin{equation}
    \delta \beta_\mathrm{Poisson} = \frac{1}{\bar{k}} \sqrt{\frac{2(1+\hat{\beta})}{ (1+M) N_\mathrm{pixel}N_\mathrm{frame}}},
    \label{equation: error_poisson}
\end{equation} 
where $M \approx 1$ characterizes the intensity fluctuations of the double pulses incident onto the sample (see Supplementary Material Section~2 for the derivation). Such a high level of agreement shows that Equation~\ref{equation: chi2} is an efficient contrast estimator and the shot noise is the main source of error even at this extremely low count rate. This agreement is non-trivial, because it suggests that the measurement accuracy of the fs-XPCS experiment is merely limited by photon statistics and thus can be only improved by increasing data volume. It is also worth noting that due to the count rate being halved, the error bars for the single-pulse contrasts are larger compared to those in the both-beam mode, despite that they are calculated using more than half a million frames as shown in Fig.~\ref{fig:fig3} (\textbf{A}). 

\begin{figure}[h!]
    \centering
    \includegraphics[width=0.85\textwidth]{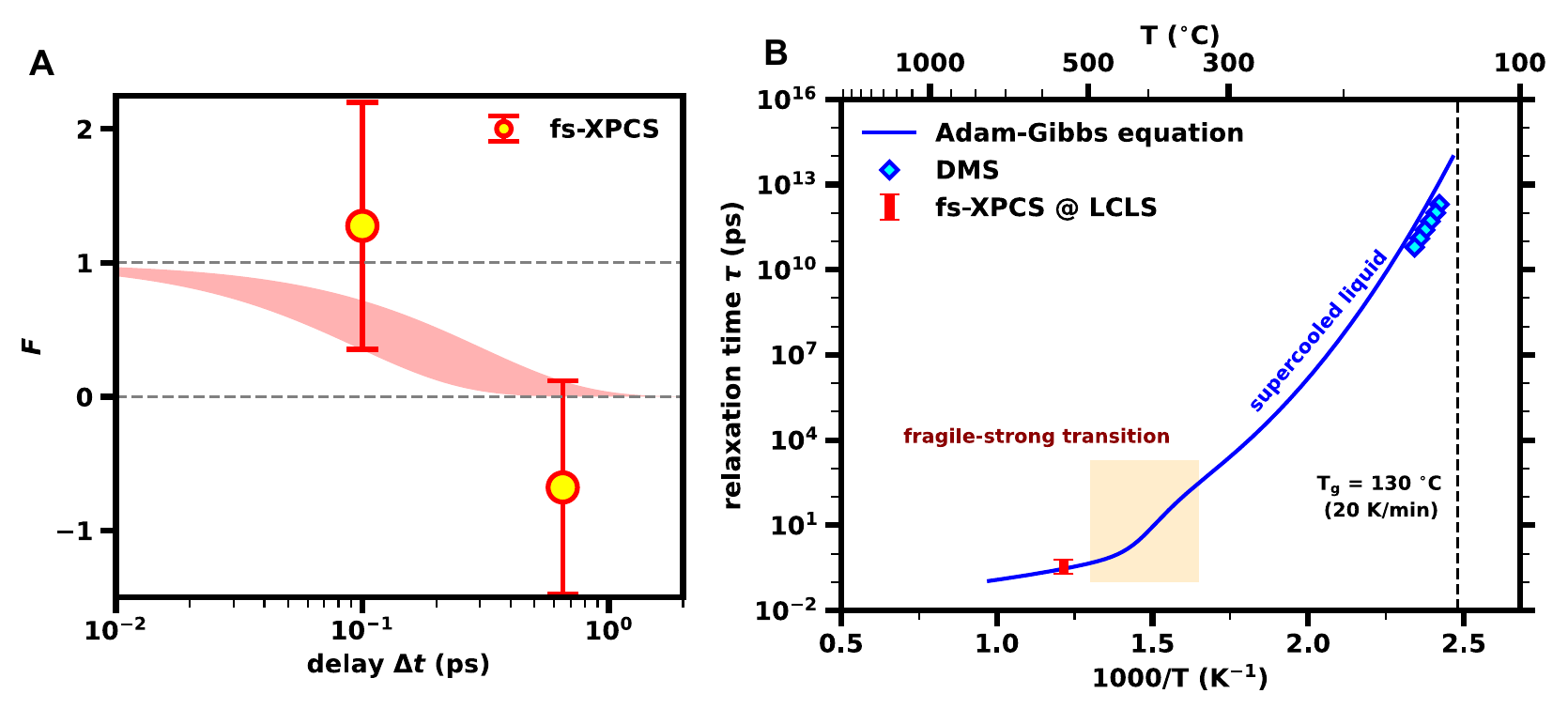}
    \caption{
    \textbf{The wide-angle fs-XPCS for determining the atomic-scale relaxation dynamics of liquid Ge$\bm{_{15}}$Te$\bm{_{85}}$.} 
  (\textbf{A}) $F$ obtained from fs-XPCS (red circles) at two time delays indicate a close to full decorrelation at 0.65 ps. The dashed lines represent the physical boundaries [0, 1.0] of $F$. The red region represents the range of all possible exponential decays that are consistent with the uncertainty margins of the fs-XPCS measurements. 
  (\textbf{B}) The boundary of $\tau$ estimated from the fs-XPCS experiment at 550~$\mathrm{{^\circ}C}$ is plotted in a red stripe and compared with the calculated $\tau \mathrm{_s}$ from the Adam-Gibbs equation (blue curve) near the FST (``double-kink" in the shaded yellow area around 400~$\mathrm{{^\circ}C}$). The eutectic melting point is about 385~$\mathrm{{^\circ}C}$. At much lower temperatures near $\mathrm{T_g}$, the temperature dependence of $\tau \mathrm{_s}$ is in agreement with that of  DMS data~\cite{peng_uncovering_2020} (diamond symbols).}
    \label{fig:fig4}
\end{figure}

With the wide-angle contrasts in the single-pulse and both-beam mode, the effective overlap coefficient $\bar{\mu}$ from the SAXS monitor, and the averaged branching ratio $\bar{r}$ from the intensity diagnostics, $F$ for each delay can be determined using Equation~\ref{eq:ISF} as $F =1.3 \pm 0.9$ for 0.1 ps and $F = -0.7 \pm 0.8$ for 0.65 ps. Since $F$ is related to the ISF $f$ via $F=|f|^2$, for $|f|^2$ having a physical meaningful value, the boundary condition $0\leq|f|^2\leq1$ must be applied. Thus, the error range of $|f|^2$ is from 0.4 to 1.0 (denoted as [0.4, 1.0]) for 0.1 ps and is from 0 to 0.1 (denoted as [0, 0.1]) for 0.65 ps.  Figure~\ref{fig:fig4} (\textbf{A}) shows the error ranges of $|f|^2$ at the two time delays. At 0.1 ps, [0.4, 1.0] indicates a significant correlation, as one would expect at this temperature. At 0.65~ps, [0, 0.1] suggests a nearly complete decorrelation. This indicates that the relaxation timescale is on the order of hundreds of femtoseconds.
We further ruled out the possibility of first pulse induced dynamics by showing that the contrast values remain consistent within the error margin when varying pulse influx (see Supplementary Material Section~3). 

To cross-check the ISF determined by fs-XPCS, we performed a quasi-elastic neutron scattering experiment (QENS)  at the Swiss spallation neutron source SINQ~\cite{jansen_focus_1997}, Paul Scherrer Institute (see Supplementary Material Section~5). QENS allows us to measure the dynamic structure factor $S(\mathrm{Q}, \omega)$ in the frequency ($\omega$) domain at the first $S(\mathrm{Q})$ peak of neutron diffractions ($\mathrm{Q_0^N} = 2.13 \pm 0.01 ~\mathrm{\AA^{-1}}$) and at the same temperature. The quasi-elastic contribution of $S(\mathrm{Q}, \omega)$, characterizing coherent dynamics, can be well described by a single Lorentzian function (see Supplementary Fig.~5). The resulting characteristic relaxation time $\tau \mathrm{^{N}} = 0.68 \pm 0.10$~ps is consistent with the drop of the ISF at 0.65~ps from the fs-XPCS. The QENS result also indicates that the ISF displays a simple exponential decay (as the Fourier transform of the Lorentizian function).  Given the exponential shape of the ISF, we plot all the possible exponential decays that are in the form of $\exp(-2t/\tau)$ and lie within the error margins of the two data points obtained by fs-XPCS, which is represented by the red region in Figure~\ref{fig:fig4} (\textbf{A}). It gives an estimation of $\tau$ in the range of 0.19 to 0.61~ps, which appears slightly smaller than, but on the same order of magnitude of, that of QENS. Note that $\tau$ and $\tau \mathrm{^{N}}$ are not necessarily the same, as the scattering cross-sections of Ge and Te differ with respect to X-rays and neutrons, and thus are weighted differently in the total scattering intensities.

Figure~\ref{fig:fig4} (\textbf{B}) shows the $\tau$ measured using fs-XPCS plotted as the red stripe in comparison with the expected values from the Adam-Gibbs fitting curve for liquid Ge$_{15}$Te$_{85}$. The blue curve represents the calculated stress relaxation time $\tau \mathrm{_s}$ from viscosity $\eta$ via the Maxwell relation, $\tau \mathrm{_s} = \eta / G_{\infty}$, where the instantaneous shear modulus $G_{\infty}$ is assumed to be a constant and set to a value of 4.5~GPa, leading to $\tau \mathrm{_s}\sim~100$~s at the standard $\mathrm{T_g}$ = 130 $\mathrm{^{\circ}C}$~\cite{angell1995formation} (a standard $\mathrm{T_g}$ is defined with calorimetric measurements at a heating rate of 20 K/min equal to the prior cooling rate). The viscosity $\eta$ is predicted from the Adam-Gibbs equation, $\eta = \eta_0 \exp (C/\mathrm{T}S_c)$ (where $C$ and $\eta_0$ are constants, and $S_c$ is the configurational entropy), based on heat capacity and high-temperature viscosity data (see Supplementary Material Section~4). A characteristic ``double-kink" in the temperature dependence of viscosity (yellow area) has been identified as the FST at around 400~$\mathrm{^{\circ}C}$ in the liquid state~\cite{wei2015phase}. Near the glass transition, the liquid exhibits a relatively strong behavior with a fragility index $m \approx 50$ (the $m$ fragility is defined as $m = d\log\eta/d(\mathrm{T_g/T})$ at $\mathrm{T_g}$~\cite{angell1995formation}), which has been verified by calorimetric studies~\cite{wei2015phase} and dynamic mechanical spectroscopy (DMS) data (diamond symbols)~\cite{peng_uncovering_2020}. Above 500~$\mathrm{^{\circ}C}$, the liquid has a weak temperature dependency of $\tau \mathrm{_s}$ and a high fragility ($m \approx 90$)~\cite{wei2015phase}, where $\tau\mathrm{_s}$ is predicted below 1~ps. The structural relaxation time $\tau$ from the fs-XPCS experiment agrees well with the projected $\tau\mathrm{_s}$, as indicated by the red stripe on the blue line. Thus, the observed sub-picosecond dynamics provide the direct atomic-level evidence of the highly fragile behavior of liquid Ge$_{15}$Te$_{85}$ above its FST temperature. 



\section{Discussion and Conclusion} 
XPCS has become a major experimental tool for understanding the atomic-level relaxation dynamics near or below $\mathrm{T_g}$ in glasses on milliseconds~\cite{alfinelli2023amorphous} or slower timescales. The XFEL-based XPCS techniques described here extends the lower boundary of the time resolution into the femtosecond regime, limited only by the pulse duration. By adjusting the optical path of the delayed-branch, the split-delay system enables studies to probe the timescales from femto- to several nanoseconds~\cite{roseker2009performance, osaka2016wavelength, zhu2017development, lu2018development, rysov2019compact}. Individual pulses separated by tens to hundreds of nanoseconds can be produced by the nanosecond double-bunch mode using accelerator techniques~\cite{deckertwo}, as demonstrated in a study of colloidal dynamics~\cite{sun2021nonuniform}. At the European XFEL, the unique time structure of intra-train MHz pulses is shown to enable XPCS to cover the timescale of $\sim\mu$s~\cite{madsen2021materials}. Corresponding to the wide range of timescales, the entire temperature range of the Angell-plot~\cite{angell1995formation} is thus covered from the ultraviscous strong liquid near the glass transition to the highly fragile liquid well above its melting point.
Unlike the frequency-domain analysis performed in inelastic scattering experiments, the direct measurements in the time-domain with XPCS contain the information revealing higher order time correlations~\cite{shpyrko2014x}. This information are intrinsic to, and revealing of, temporally heterogeneous dynamics in disordered systems~\cite{Madsen_Leheny_Guo_Sprung_Czakkel_2010, Evenson_Ruta_2015}. A recent work by Böhmer et al.~\cite{bohmer2024time}, using multispeckle dynamic light-scattering measurements, demonstrated an experimental determination of ``material time" (a concept coined to describe the intrinsic time measured on a clock whose rate changes with glass ageing) in glass formers near the glass transition. The authors found the time reversibility of materials during ageing on the timescale of hundreds of seconds, where scattering intensity fluctuations are statistically time-reversible with the material time. To access the material time, time-domain experiments are preferred~\cite{bohmer2024time}, as it can in principle access instantaneous autocorrelations even when ageing or decorrelation takes place on comparable timescales that may change material properties. Thus, XFEL-based XPCS promises to be a crucial technique to determine if material-time reversibility may apply in fast ageing systems (e.g., an undercooled liquid just below the melting point). 

In these contexts, the XFEL-based fs-XPCS using separated pulse pairs may open a new avenue for studying the atomic-scale structural relaxations in uncharted territories (obscured by extremely short-time relaxation or fast crystallization) in the glass and liquid sciences. For instance, one could implement a laser-pulse pump shortly before the double X-ray pulses and probe the dynamics during the rapid melt-and-quench process of a liquid before fast crystallization interferes. This necessitates a high cooling rate, achievable only with the small sample sizes, as permitted by X-ray scattering. Such a pump-and-double-probe scheme might be used to access the relaxation dynamics of supercooled water below its homogeneous nucleation limit 231~K~\cite{Kanno_Angell_1979} (termed as ``no-man"s land), where the origin of the well-known thermodynamic anomalies has been debated for decades~\cite{Kim_Amann-Winkel}. Poor glass forming phase-change materials (e.g. $\mathrm{Ge_{15}Sb_{85}}$ and others~\cite{zalden2019femtosecond,wei2019phase}) undergo a metal-to-semiconductor fragile-strong transition during supercooling (20-30\% below $\mathrm{T_m}$) before they crystallize within a few nanoseconds. Measuring the atomic-scale relaxation dynamics near these ``short-lived" transitions is of particular relevance for understanding the functionality of these materials in phase-change memory devices. Although inelastic neutron scattering (including QENS) is conventionally used to measure dynamics on the energy transfer of meV to $\mathrm{\mu}$eV, corresponding to timescales from picoseconds to nanoseconds, these techniques require a large sample and beam size, on the order of one to several centimeter, due to weak interaction of neutrons with matter and the limited neutron flux. This makes it difficult to achieve substantial undercooling in liquids that are poor glass formers, or produce a large quantity of samples. In addition, neutron incoherent scattering occurs usually on the meV energy transfer, interfering the signal of interest for coherent atomic dynamics~\cite{Alfred Chapter}. When applying the Fourier transform to frequency-domain data, the accuracy is affected by the cut-off frequency and choice of data range to be transformed, leading to potential uncertainties.  As an alternative frequency-domain technique, inelastic X-ray scattering (IXS) can probe small samples with a focused X-ray beam ($\sim$10 to 100~$\mu$m beam spot size) (~\cite{Alfred Chapter}). Owing to the high energy of X-rays of order $\sim$10~keV, the energy resolution of IXS is limited and thus is suited to probe large energy transfer of scattering (e.g. phonon dynamics). Yet resolving meV energy transfer requires extremely high energy resolution $\Delta E/E\sim10^{-7}$~\cite{Alfred Chapter}. Recent progress of high-resolution IXS makes it possible to probe 1 to 100~meV; however, resolving $\sim1$~meV or below (i.e. order of $\sim 1$~ps or larger) remains increasingly challenging~\cite{Alfred Chapter}.

It is worth noting that approximately 1.7 million frames are analyzed in this study to determine the ISF at two delay points. The number of delay points is constrained by the pulse repetition rate of 120 Hz presently available at the LCLS, and the precision of our measurements is primarily limited by shot noise.  
Thus, data collection for each delay point requires several hours. However, this scenario will undergo a transformative change with the advent of the LCLS-II-HE, which will provide hard X-ray pulses with MHz repetition rates~\cite{raubenheimer2018lcls}. 
Together with the cutting-edge multi-mega-pixel X-ray detectors such as CITIUS~\cite{nishino2023false}, SParkPix-S~\cite{SparkPix-ED} and AGIPD~\cite{allahgholi2019adaptive}, which operate at tens of kHz to MHz frame rates, it naturally allows us to collect data at substantially higher rates ($\sim$ four orders of magnitude faster), potentially reducing the measurement time per delay point to tens of seconds. 
The methodology demonstrated in this work can then be efficiently employed to investigate the structural relaxation over the entire temperature range where the dynamic crossovers may occur.

\section*{Acknowledgement}
Use of the Linac Coherent Light Source (LCLS), SLAC National Accelerator Laboratory, is supported by the U.S. Department of Energy, Office of Science, Office of Basic Energy Sciences under Contract No. DE-AC02-76SF00515. The QENS results are based on experiments performed at the Swiss spallation neutron source SINQ, Paul Scherrer Institute, Villigen, Switzerland. Y.S. would like to acknowledge many inspiring discussions with Mark Sutton, from experimental design to data analysis. S.W. and Y.S. thank Peter Zalden for his helpful comments on the manuscript. This work was supported by a research grant (42116) from VILLUM FONDEN (S.W., T.F.) and partially supported by Deutsche Forschungsgemeinschaft (DFG, German Research Foundation) (WE 6440/1-1). B.R. and A.C. acknowledge funding for this project from the European Research Council (ERC) under the European Union’s Horizon 2020 research and innovation programme (Grant Agreement No 948780). We thank the Danish Agency for Science, Technology, and Innovation for funding the instrument center DanScatt. K.S.T. and T.J.A. acknowledges financial support by the DFG through Project No. 278162697-SFB 1242. P.L. acknowledges financial support from NSF-DMR under Grant No. 1832817. Work of H.L. was supported by U.S. Department of Energy, Office of Science under DOE (BES) Awards DE-SC0022222.

\subsection*{Author contributions}
YS and SW conceived and led the project. YS, HL, TJA, SS, TS, KST, DZ, JM, AC, PS, SW, MM, CY, JVC, VE, JV, NW, JH, TM, CD, PF designed, tested, and performed the fs-XPCS experiment with samples from NA and JM, and with furnace by YC, YS, and FY; TF and YS analyzed the fs-XPCS data. SW, TF, PS, JM, AM, and JE performed the QENS with samples from GC and PL. The QENS data were analyzed by TF and SW with input from JPE and FY. YS, SW, and TF wrote the paper with important input from BR, PF, DZ, FY, TJA, AC, KST, and other authors.
\bibliography{scibib}

\bibliographystyle{Science}

\end{document}


\maketitle 
\textbf{This PDF file includes:} \\
 \\
Supplementary text 
 \begin{enumerate}
         \item Intensity distribution of the double pulses from the split-delay system
         \item Contrast calibration and extraction
         \item Evaluation of the X-ray beam perturbation
         \item The prediction curve based on the Adam-Gibbs equation
         \item Quasi-Elastic Neutron Scattering measurement
\end{enumerate}
Figures 1-5 (Supplemental) \\


\section{Intensity distribution of the double pulses from the split-delay system}
\label{section: intensity_distribution}
\begin{figure}[H]
    \centering
    \includegraphics[width=1\linewidth]{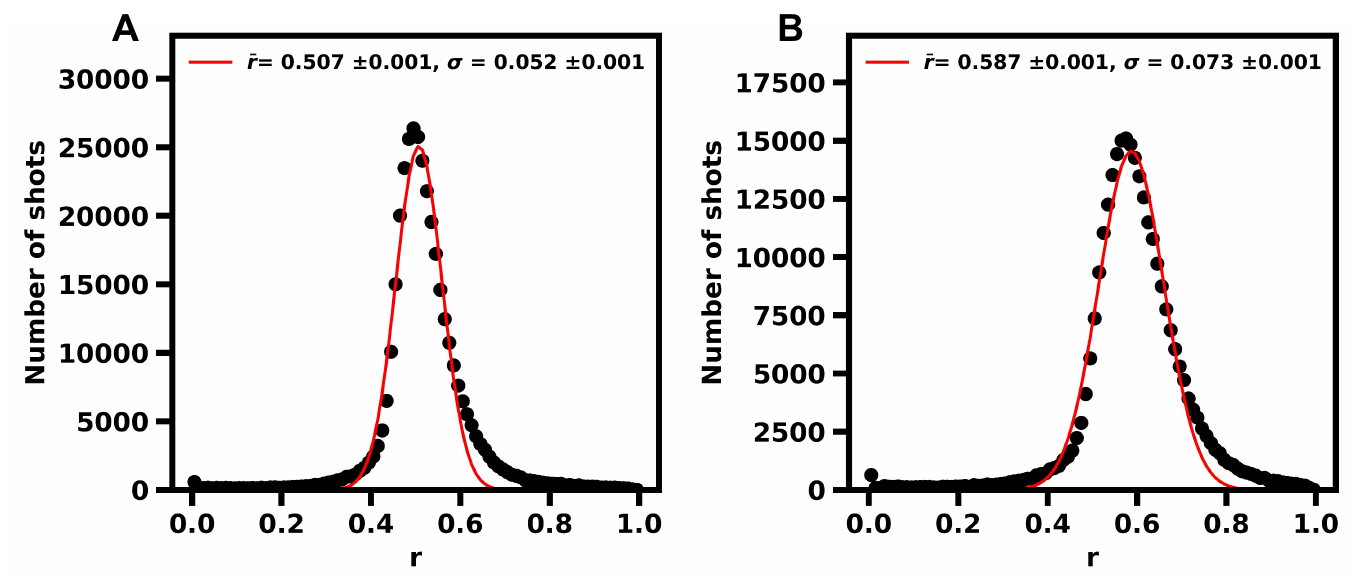}
    \caption{(Supplemental) \textbf{Histograms of the branching ratio $\bm{r}$ of the split-delay system during the WAXS measurements}. (\textbf{A}) Histogram of the branching ratio for the double pulses at a delay of 0.1~ps. (\textbf{B}) Histogram of the branching ratio for the double pulses at a delay of 0.65~ps . Red lines are the Gaussian fits giving the averaged value and the RMS spread of the branching ratio.}
    \label{fig:intensity_analysis}
\end{figure}
As mentioned in the manuscript, transmissive intensity diagnostics consisting of a  PIPS (passivated implanted planar silicon; Canberra PIPS FD300) diode gathering the scattering from a Kapton target are placed in the optical path of each branch of the split-delay system to measure their intensities, i.e., $i_\mathrm{fix-delay}$ and $i_{\mathrm{delayed}}$. This allows the analysis of the distribution of the shot-to-shot branching ratio $r = i_\mathrm{fix-delay}/(i_\mathrm{fix-delay}+i_\mathrm{delayed})$, which is displayed in Fig.~\ref{fig:intensity_analysis}. By fitting to a Gaussian distribution, the averaged branching ratio as well as its spread can be retrieved. They are also listed in Table~1 and used to determine the ISF in the main paper. 

\section{Contrast calibration and extraction}
\label{section: contrast_calibration}
\begin{figure}[H]
    \centering
    \includegraphics[width=0.75\linewidth]{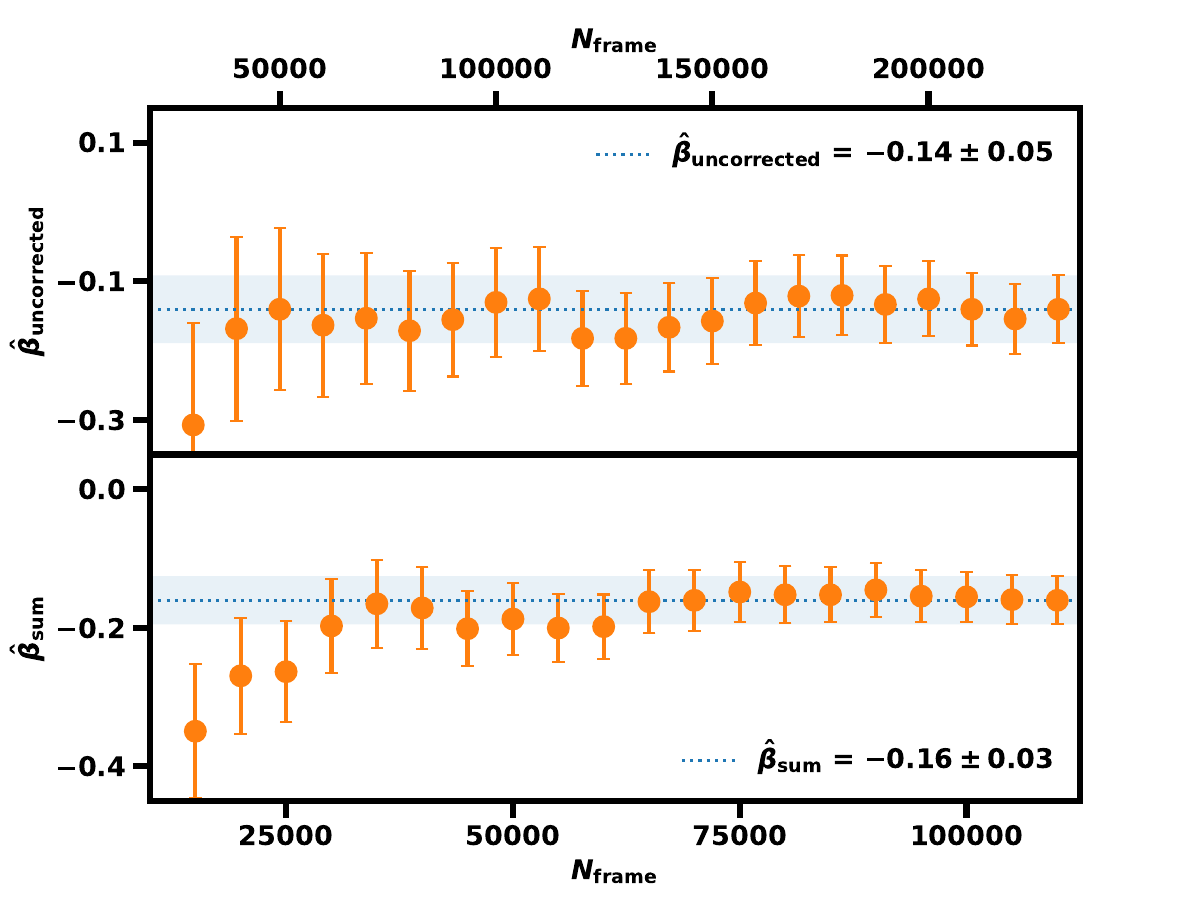}
    \caption{(Supplemental) \textbf{Convergence and calibration of the speckle contrast extraction.} \textbf{Upper panel}: raw/uncorrected contrast directly from the greedy guess photon assignment algorithm as a function of the number of frames used in the calculation. The data is measured with both beams on sample having a time delay of 0.65~ps. \textbf{Lower panel}: Contrast convergence from the artificial scattering sum using the same frames. The dotted line and the blue shade show the final contrast value and its MLE error.}
    \label{fig:contrast_calibration}
\end{figure}

As described in Ref.~\cite{hruszkewycz2012high}, at the discrete photon limit, the intensity distribution of a speckle pattern, i.e., the probability of $k$ photon(s) per pixel events $P_k$ follows negative binomial distribution~\cite{Mandel_Wolf_book}
\begin{equation}
       P_k(\beta,\bar{k}) = \frac{\Gamma(k+{1/\beta)}}{\Gamma(1/\beta) \Gamma(k+1)} [(\frac{\bar{k}\beta}{\bar{k}\beta+1})^k(\frac{1}{\bar{k}\beta+1})^{1/\beta}].
\end{equation}
Here $\bar{k}$ is the count rate. The $k$-photon probabilities are directly determined by the contrast $\beta$. For example, since the count rate is very low ($\bar{k} \ll 1$ photons per pixel), Taylor expansion of the one and two-photon probability gives
\begin{equation}
       P_1 = \bar{k} - (1+\beta)\bar{k}^2 + O(\bar{k}^3),
    \label{eq:P1} 
\end{equation}

\begin{equation}
    P_2 = \frac{1}{2} (1+\beta) \bar{k}^2 + O(\bar{k}^3)
    \label{eq:P2}
\end{equation}
Contrast extraction therefore relies on comparing the measured $k$-photon probability $p_k$ with the theoretical $P_k$. Taking the derivative of Equation~\ref{eq:P2}, we have
\begin{equation}
     \frac{\delta P(2)}{P(2)} = \frac{\delta \beta}{1+\beta}
\end{equation}
Considering the signal-to-noise ratio of photon counting:
\begin{equation}
    \frac{\delta P(2)}{P(2)} =  \frac{1}{\sqrt{  \frac{1}{2}
 \sum_{f=1}^{N_{\mathrm{frame}}} (1+\beta) N_{\mathrm{pixel}}\bar{k_f}^2}}.
\end{equation}
With constant X-ray flux corresponding to a count rate $\bar{k}$, we have
\begin{equation}
   \delta \beta = \frac{\sqrt{2(1+\beta)}}{\bar{k}\sqrt{N_{\mathrm{frame}} N_{\mathrm{pixel}}}}
   \label{eq:constant_intensity_error}
\end{equation}
With monochromatic SASE pulses, the intensity fluctuates shot-to-shot, assume
\begin{equation}
   \frac{1}{N_{\mathrm{frame}}} \sum_{f=1}^{N_{\mathrm{frame}}} \bar{k_f}^2 = (1+M) (\frac{1}{N_{\mathrm{frame}}} \sum_{f=1} \bar{k_f})^2= (1+M) \bar{k}^2
   \label{eq:fluctuating_intensity}
\end{equation}
We therefore have the error estimation 
    \begin{equation}
   \delta \beta_\mathrm{Poisson} = \frac{\sqrt{2(1+\beta)}}{\bar{k}\sqrt{(1+M) N_{\mathrm{frame}} N_{\mathrm{pixel}}}}
   \label{eq:fluctuating_intensity_error}
\end{equation}
The intensity monitors mentioned in Section~\ref{section: intensity_distribution} in the Supplementary Material measure the shot-to-shot intensity and yield $M\approx 1$. 

Plotted in Fig.~\ref{fig:contrast_calibration} upper panel is the raw contrast extracted from the WAXS at a delay of 0.65~ps. When increasing the number of frames,  the extracted contrast value converges to a negative value with $\hat{\beta}_\mathrm{uncorrected} = -0.14 \pm 0.05$. 
This systematic bias in contrast estimation has been analyzed in the previous simulation and experimental studies: the charge sharing between neighboring pixels of hard X-ray detectors leads to unavoidable systematic errors in the photon assignment of the commonly used algorithms including the greedy guess method used in this study~\cite{sun2020accurate, sun2021contrast}. 
Nevertheless, the error can be properly accounted for by applying a linear correction to get the calibrated contrast $\hat{\beta}_\mathrm{cal}$
\begin{equation}
   \hat{\beta}_{\mathrm{cal}} = \frac{\hat{\beta}_{\mathrm{uncorrected}}+1}{\alpha} - 1
\end{equation}
Here the correction factor $\alpha$ can be retrieved through a procedure of making artificial scattering sums. Since individual detector frames have a time spacing of multiples of 8.3~ms, which corresponds to the 120 Hz detector frame rate and repetition rate of the X-ray source. This is much longer than the decorrelation timescale of the sample system. As a result, by randomly picking and adding the two frames of equal intensity, the summed scattering should have a contrast half of that from single frame. Given the extracted contrast of the artificial sum to be 
$\hat{\beta_{\mathrm{sum}}}$, the correction factor can be derived
\begin{equation}
\alpha = 2\hat{\beta}_{\mathrm{sum}} 
+1 - \hat{\beta}_{\mathrm{uncorrected}}
\end{equation}
The equal-intensity artificial  sum is calculated with the same data as used in the upper panel, and the contrast value converges to
 $\hat{\beta}_{\mathrm{sum}} = {-0.16} \pm {0.03}$. The correction factor is therefore determined to be $\alpha = {0.82} \pm {0.08}$.
\begin{figure}[H]
    \centering
    \includegraphics[width=0.65\linewidth]{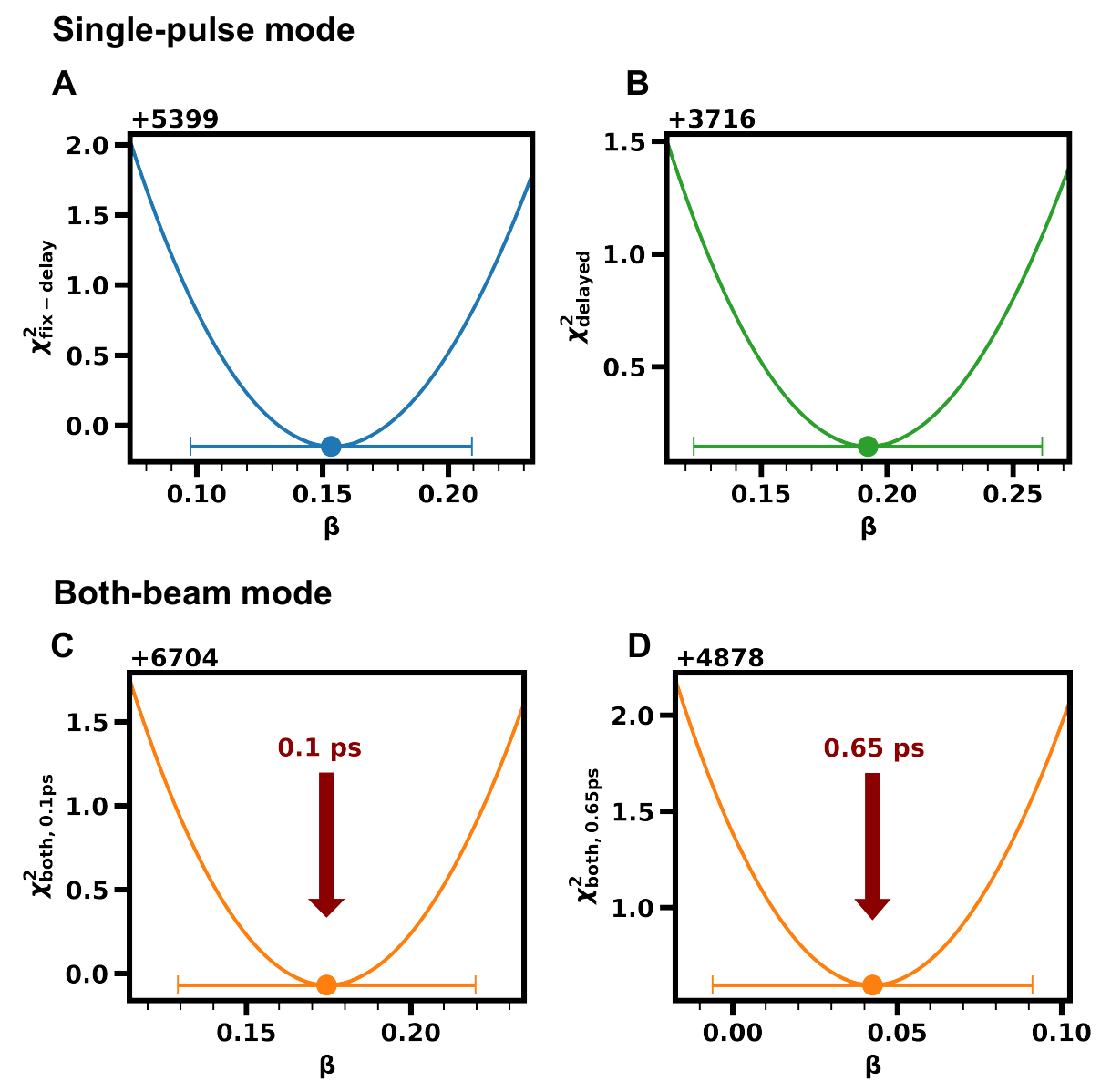}
    \caption{(Supplemental) \textbf{Maximum likelihood estimator (MLE) for contrast extraction.} The contrast values $\hat{\beta}$ are determined using numerical calculations of $\chi^2$ as a function of $\beta$. The minimized $\chi^2$ yields the best estimate, i.e., $\hat{\beta}$ for \textbf{A}) the fixed-delay branch, \textbf{B}) the delayed branch, \textbf{C}) the both-branch for 0.1 ps, and \textbf{D}) the both-branch for 0.65 ps. Clearly, the contrast for the both-branch decreases substantially from $0.17~\pm ~0.05$ at 0.1 ps (\textbf{C}) to $0.04~\pm~0.05$ at 0.65 ps (\textbf{D}) (as tabulated in Table~1 in the main paper).}
    \label{fig:MLE_estimator}
\end{figure}
Since droplet based photon assignment algorithms including the greedy guess method is accurate in determining number of photons (or the count rate $\bar{k}$)~\cite{sun2020accurate}, the correction factor can be understood as an underestimation of the two photon probability $p_2$ and an overestimation of the one photon probability $p_1$, which can be calibrated
\begin{equation}
    p_{\mathrm{1, cal}} =   p_1 - 2(\frac{p_2}{\alpha} - p_2);   \quad  p_{\mathrm{2, cal}} = \frac{p_2}{\alpha}
\end{equation}
The calibrated probabilities are subsequently used for contrast extraction using the maximum likelihood estimator for all WAXS measurements, including the two single-pulse modes, (i.e., fix-delay and delayed branch) as well as the both-beam mode at 0.1~ps and 0.65~ps, as shown in Fig.~\ref{fig:MLE_estimator} (\textbf{A}-\textbf{D}). 

\section{Evaluation of the X-ray beam perturbation}
\label{section: beam_effect}
\begin{figure}[H]
    \centering
    \includegraphics[width=0.65\linewidth]{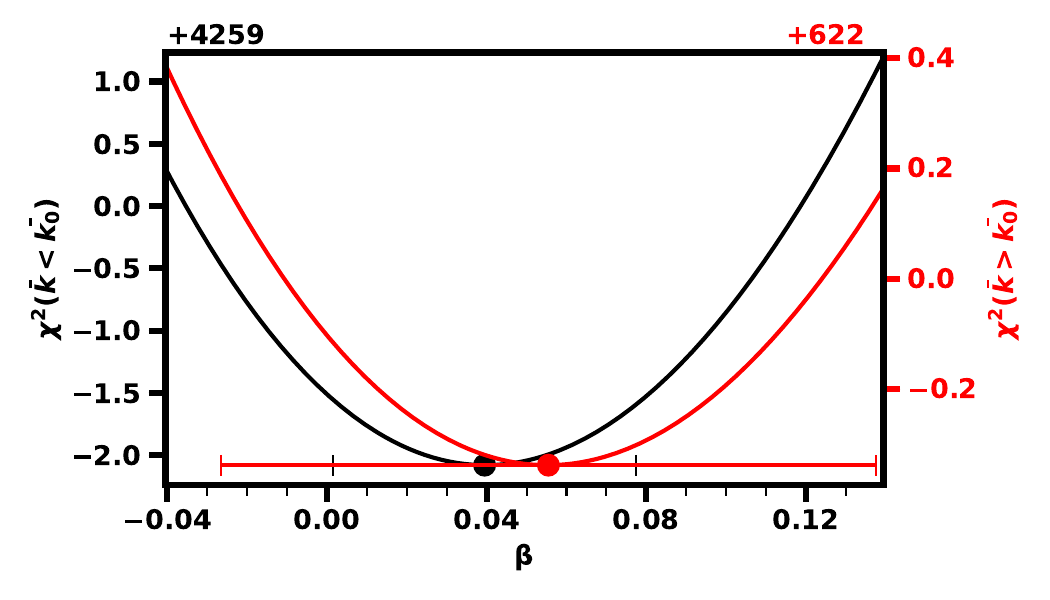}
    \caption{(Supplemental) \textbf{Evaluation of the beam perturbation effect.} At a delay of 0.65~ps, the MLE method is applied towards the scattering data grouped by the count rate with a threshold value $\bar{k_0} = 2\times 10^{-4}$~photons/pixel. Although the low (black) and high (red) count rates differ about sixfold, the contrasts of the two groups show a good agreement  within the error range, indicating no pulse-induced contrast change.}
    \label{fig:beam_effect}
\end{figure}
The possibility of first pulse induced dynamics can be ruled out by the contrast analysis with varying pulse influx. Since the pulse influx is approximately proportional to the count rate, the scattering data for the delay of 0.65~ps are grouped with respect to the count rate by a threshold value $\bar{k_0} = 2\times 10^{-4}$~photons/pixel. The contrast determination of the low and high intensity groups are plotted as black and red in Fig.~\ref{fig:beam_effect}. Between the two groups, the average incident pulse flux differs approximately sixfold, but the contrast values of the two groups show little change within the error margin. This indicates that the reduction in contrast observed at the delay of 0.65~ps arises from the structural relaxation of the sample, not related to the X-ray pulse itself.

\section{The prediction curve based on the Adam-Gibbs equation}
\label{section:AG}

The predicted relaxation times in Fig.~4 (\textbf{B}) (blue curve) in the main paper is based on the fit with the Adam-Gibbs equation to the heat capacity and high-temperature viscosity data. The details of the fit are documented in Ref.~\cite{wei2015phase}. The results are the predicted viscosity values over 15 orders of magnitude (Fig.~9 of Ref.~\cite{wei2015phase}). Since viscosity $\eta$ is assumed to be proportional to the stress relaxation time $\tau\mathrm{_s}$, which is the average timescale of stress relaxation in response to shear strain. $\tau\mathrm{_s}$ is calculated using $\tau\mathrm{_s} = \eta / G_{\infty}$, where $G_{\infty}$ is the instantaneous shear modulus and assumed as a constant over the entire temperature range. $G_{\infty}$ is set to a value of 4.5 GPa which leads to $\tau\mathrm{_s} \sim 100$~s at the standard glass transition 130 $\mathrm{^{\circ}C}$. With this $G_{\infty}$, $\tau\mathrm{_s}$ can be calculated over the entire temperature range and plotted in Fig.~4 (\textbf{B}) in the main paper. Earlier studies indicated that at the standard glass transition, relaxation times for most glasses approximate $\sim 100$ s and their viscosity approaches $\sim 10^{12}$ Pa~s, although these values may differ across various systems, and even within the same system, the values may exhibit discrepancies depending on the specific relaxation processes being measured such as stress, enthalpy, dielectrics, and structures observed on different length scales~\cite{angell1991relaxation, hodge1994enthalpy, moynihan1976structural, monnier2020vitrification}.

\section{Quasi-Elastic Neutron Scattering measurement}
\label{section:QENS}
\begin{figure}[H]
    \centering
    \includegraphics[width=1\linewidth]{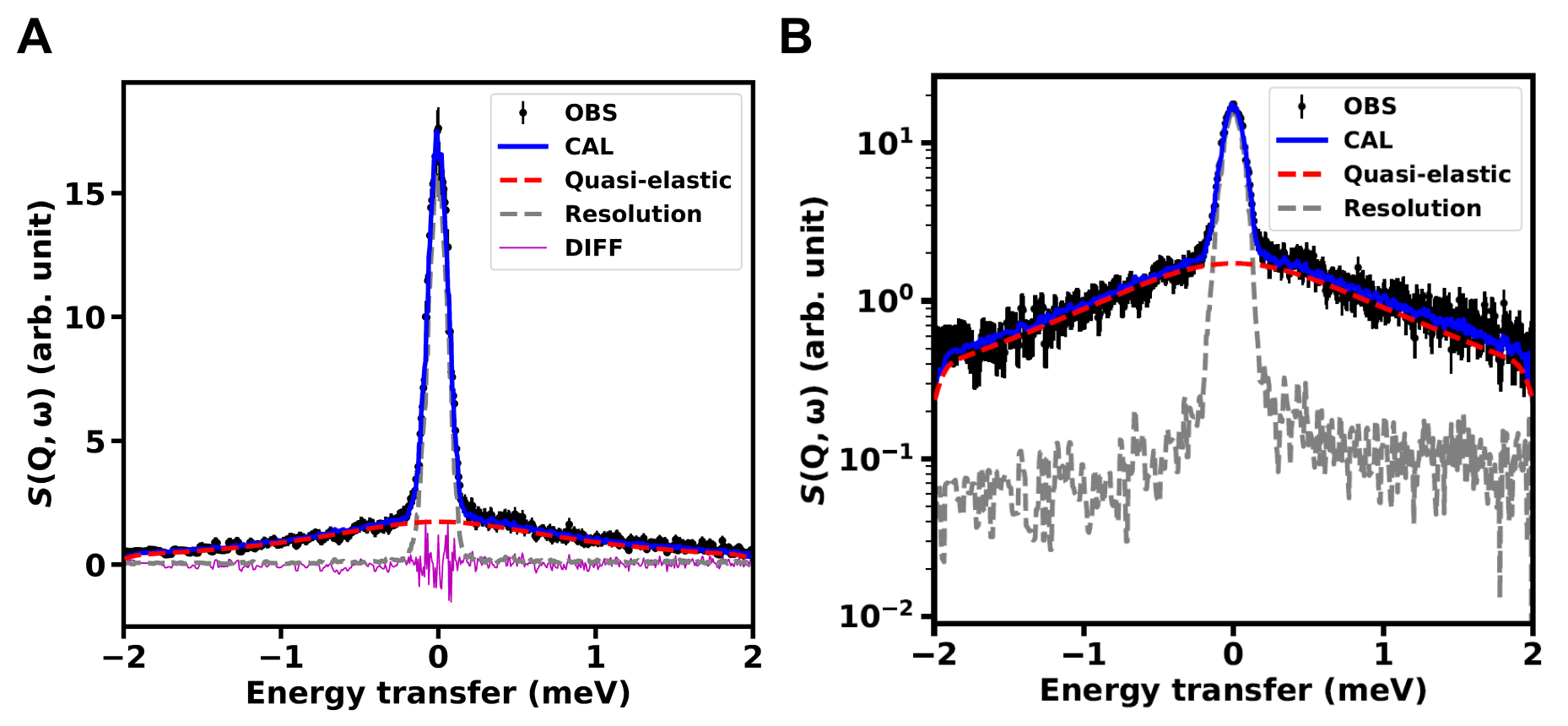}
    \caption{(Supplemental) \textbf{Profile fitting of the QENS experiment on Ge$\bm{_{15}}$Te$\bm{_{85}}$ at 550~$\bm{\mathrm{^{\circ}C}}$} in \textbf{A)} linear and in \textbf{B)} logarithmic scale. The fitting was performed at $\mathrm{Q_0^N} \approx 2.13$~$\mathrm{\AA^{-1}}$.  The red dashed line gives the quasi-elastic contributions expressed by a single Lorentzian function convoluted with the energy resolution function. The dashed gray line is the elastic contribution ($\delta$-function) convoluted with the energy resolution function. The blue line is the sum of all the contributions of the fitting model. The nearly zero values of the DIFF line in (\textbf{A}), representing the difference between the measured data and the fitted model, indicate a good fit of the model.}
    \label{fig:qens}
\end{figure}

Quasi-elastic neutron scattering (QENS) measurements were performed at Swiss Spallation Neutron Source SINQ of Paul-Scherrer Institute, using the time-of-flight (TOF) spectrometer FOCUS~\cite{jansen_focus_1997}. The sample Ge$_{15}$Te$_{85}$ was sealed in a quartz tube with the inner diameter of 8 mm and the wall thickness of 1~mm. The quartz tube was loaded into an Al$_{2}$O$_{3}$ crucible, which was subsequently placed into the furnace with the Nb sample holder. The sample was heated up at a heating rate of 10~K/min to 550~$\mathrm{^{\circ}C}$ and held isothermally for QENS measurements. The wavelength of the incident cold neutrons was selected as $\lambda =$ 4.4~$\mathrm{\AA}$, yielding an accessible momentum transfer range $0.3 < \mathrm{Q} < 2.5$~$\mathrm{\AA^{-1}}$ at zero energy transfer. The normalization and the correction for the energy dependent detector efficiency were performed on the software DAVE~\cite{dave2009}. The TOF data were calibrated with the vanadium standard and converted to the dynamic structure factor $S(\mathrm{Q}, \omega)$. The spectrum is found to be well described by the model,

\begin{equation}
    S(\mathrm{Q},\omega)=R(\mathrm{Q}, \omega)\otimes N[A_0\delta(\omega)+(1-A_0)L(\mathrm{Q}, \omega)] +b(\mathrm{Q}, \omega),
\end{equation}
where $R(\mathrm{Q}, \omega)$ is the instrumental resolution function, $\otimes$ denotes a numerical convolution, $N$ is a normalization factor, $L(\mathrm{Q},\omega)$ represents the quasi-elastic scattering from the sample melt, $\delta(\omega)$ function is the elastic scattering from the container, $A_0$ is the ratio of elastic to total scattering intensity, and $b(\mathrm{Q}, \omega)$ is a flat background and fixed to zero due to its negligibly small contribution. The profile fitting was performed for the corrected data on the same software, as plotted in Fig.~\ref{fig:qens}. The quasi-elastic broadening is found to be best described with a single Lorentzian of the form,
\begin{equation}
    L(\mathrm{Q},\omega)=\frac{1}{\pi}\frac{\Gamma(\mathrm{Q})}{(\hbar\omega)^{2}+\Gamma(\mathrm{Q})^{2}}
\end{equation}
where $\Gamma$ is the half-width at half-maximum (HWHM). The profile fitting yields the relaxation time $\tau\mathrm{^N} = 0.68~\pm~0.1$~ps  via $\Gamma = \hbar/\tau\mathrm{^N}$ at the first peak position of neutron diffraction $\mathrm{Q_0^N} \approx 2.13~\pm~0.01~\mathrm{Å^{-1}}$~\cite{kaban2007temperature}. Note that $\mathrm{Q_0^N}$ is slightly larger than $\mathrm{Q_0}=2.0~\mathrm{Å^{-1}}$ for X-ray scattering. The error range of  $\tau\mathrm{^N}$ is estimated by a sum of the errors of fitting, the selection of possible fitting ranges in available energy transfer from -3.5 to 2.0~meV, and the uncertainty of $\mathrm{Q_0^N}$. The peak positions of $\delta(\omega)$ function and the Lorentzian distribution were fixed at zero.

\bibliography{scibib}
\bibliographystyle{Science}